\documentclass[12pt]{article}

\pdfoutput=1

\usepackage[top=80pt,bottom=85pt,left=85pt,right=85pt]{geometry}
\usepackage{amsmath,amssymb,graphicx,float,color,tikz,cite,savesym,wasysym}
\usepackage{caption}
\usepackage{subcaption}
\usepackage[debug,pageanchor=false]{hyperref}
\definecolor{link}{rgb}{.8,.15,.1}
\hypersetup{colorlinks=true,linkcolor=link,citecolor=link,urlcolor=link,linktocpage}

\makeatletter
\@addtoreset{equation}{section}
\makeatother
\savesymbol{iint}
\savesymbol{iiint}

\restoresymbol{TXF}{iint}
\restoresymbol{TXF}{iiint}

\begin{document}

	\begin{titlepage}

	\begin{center}

	\vskip .5in %.3in 
	\noindent

	{\Large \bf{AdS$_8$ Solutions in Type II Supergravity}}

	\bigskip\medskip

	 Clay C\'{o}rdova,$^1$ G.~Bruno De Luca,$^{2}$ Alessandro Tomasiello$^{2}$\\

	\bigskip\medskip
	{\small 
$^1$ School of Natural Sciences, Institute for Advanced Study, Princeton, NJ 08540, USA
\\	
	\vspace{.3cm}%{.1cm}
	$^2$ Dipartimento di Fisica, Universit\`a di Milano--Bicocca, \\ Piazza della Scienza 3, I-20126 Milano, Italy \\ and \\ INFN, sezione di Milano--Bicocca
	
		}

	\vskip .5cm %.3cm
	{\small \tt claycordova@ias.edu, g.deluca8@campus.unimib.it,  alessandro.tomasiello@unimib.it}
	\vskip .9cm %.6cm
	     	{\bf Abstract }
	\vskip .1in
	\end{center}

	\noindent

	We find non-supersymmetric AdS$_8$ solutions of type IIA supergravity. The internal space is topologically an $S^2$ with a U(1) isometry. The only non-zero flux is $F_0$; an O8 sourcing it is present at the equator of the $S^2$. The warping function and dilaton are non-constant. It is also possible to add D8-branes on top of the O8. Possible destabilizing brane bubbles (whose presence would be suggested by the weak-gravity conjecture) are either absent or collapsing.  Our solutions are candidate holographic duals to unitary interacting CFTs in seven dimensions with exceptional global symmetry.  We also present analogous non-supersymmetric AdS$_{d}$ solutions for general $d$ which are supported only by $F_0$.

	\noindent

	\vfill
	\eject

	\end{titlepage}

\tableofcontents
	
\section{Introduction} % (fold)
\label{sec:intro}

Quantum field theory is a universal framework for describing the behavior of many physical phenomena.  It is useful to organize the dynamics by energy scale via the renormalization group. At the shortest and longest distances one frequently finds conformally invariant systems, and thus CFTs play a foundational role in our understanding of field theory.

In simple examples in low spacetime dimensions, models of interacting CFTs can be found starting from free fields and tuning interactions to a critical point.  In spacetime dimension $d>4$ this simple paradigm breaks down, since all interactions of free fields are either unstable or irrelevant.  This makes the problem of defining ultraviolet complete interacting theories in high spacetime dimensions challenging.  Nevertheless, string theory sometimes suggests the existence of critical points engineered by intersecting branes \cite{Witten:1995zh, Strominger:1995ac, Witten:1995em, seiberg-5d}.  Recent years have seen a revival of interest in such CFTs, fueled for example by progress in their holographic AdS$_{d+1}$ duals (see for example \cite{afrt,10letter,cremonesi-t} for $d=6$ and \cite{dhoker-gutperle-karch-uhlemann,dhoker-gutperle-uhlemann} for $d=5$), by F-theory \cite{heckman-morrison-vafa,heckman-morrison-rudelius-vafa}, or by field-theoretic analysis \cite{osty-a6, intriligator-a6, cordova-dumitrescu-yin, cordova-dumitrescu-intriligator-a6 }. 

The concrete examples of $d>4$ CFTs that emerge from string theory are all supersymmetric, and this enhanced symmetry provides crucial insights to their dynamics.  However, for algebraic reasons, superconformal field theories can only exist in $d\leq 6$ \cite{nahm,minwalla}.  Thus we are left to wonder whether interacting CFTs can exist in general spacetime dimensions.  In other words: does unitary quantum field theory itself have an upper critical dimension, i.e.\ a dimension beyond which all unitary theories are necessarily free?

In this paper we confront this problem via gauge-gravity duality.  We construct new non-supersymmetric solutions of type IIA supergravity where the only non-vanishing flux is $F_{0}$.  In particular, these solutions can have AdS$_{8}$ factors and hence are potentially holographically dual to $d=7$ interacting CFTs.  Note that while any effective theory in AdS$_{d+1}$ defines a perturbative solution to the crossing equations of a putative dual CFT$_{d}$ \cite{heemskerk-penedones-polchinski-sully}, the embedding in string theory strongly suggests that our models are non-perturbatively consistent.\footnote{Another approach to CFTs in high dimensions is the numerical bootstrap. See \cite{fitzpatrick-kaplan-poland} for preliminary discussion.  See also  \cite{giombi-perlmutter} for another recent proposal of non-supersymmetric holography.}  

Apart from CFT motivations, the study of high-$d$ compactifications is also interesting as a simple version of the landscape problem. Indeed supersymmetric AdS$_7$ and AdS$_6$ solutions have by now been completely classified (see \cite{afrt,10letter,cremonesi-t} and \cite{dhoker-gutperle-karch-uhlemann,dhoker-gutperle-uhlemann}), and one can hope that this will inspire progress in the harder classification of $d=4$ compactifications. We can thus view AdS$_8$ as a simple setup where the restricted geometry might enable a classification of non-supersymmetric compactifications, parallel to the classification of supersymmetric AdS$_7$ compactifications. 

Our strategy for finding AdS$_8$ solutions is straightforward. We assume that the internal $M_2$ space has a U(1) isometry. This reduces the equations of motion to a system of ODEs. We study these equations of motion first in a perturbation series, and then numerically. The perturbation approach is especially useful to treat loci where the isometry $S^1$ shrinks. We present this analysis in section \ref{sec:o8-d8} below.

\begin{figure}
\centering
\begin{subfigure}{.5\textwidth}
  \centering
  \includegraphics[width=.4\linewidth,  height=0.5\textwidth]{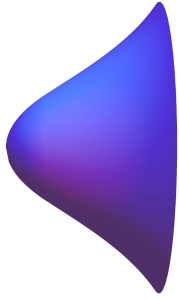}
  \caption{\small The O8 Solution.}
  \label{fig:O8}\label{eq:}
\end{subfigure}%
\begin{subfigure}{.5\textwidth}
  \centering
  \includegraphics[width=.4\linewidth,  height=0.5\textwidth]{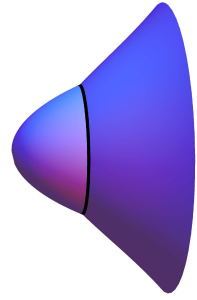}
  \caption{\small A Solution with D8s away from the O8.}
  \label{fig:O8D8}
\end{subfigure}
\caption{\small Cartoons of the compactification geometry $M_{2}$ for our solutions. In (\subref{fig:O8}), the solution with no D8-branes: the radius of the circle and string coupling diverge as we approach the O8 brane where the geometry ends.  In (\subref{fig:O8D8}) an example with D8-branes (shown in black) away from the O8.  As discussed in section \ref{sub:stab}, these solutions are unstable: the D8s either slip off to the tip of the geometry or to the O8.}
\label{fig:test}
\end{figure}

One class of solutions that emerges with this treatment has an $M_2$ which is topologically an $S^2$, with an O8-plane with infinite string coupling at its equator. This makes it similar to existing AdS$_6$ \cite{brandhuber-oz}, AdS$_7$ \cite[Sec.~5.1]{bah-passias-t} solutions.\footnote{The AdS$_3$ solutions in \cite{dibitetto-lomonaco-passias-petri-t} also have an O8 of the same type, but have a non-compact internal space.} What makes it far simpler than those, however, is that the only flux present is the Romans mass $F_0$.  A cartoon of this geometry is shown in figure \ref{fig:O8}.  We can also generalize this class of examples by including D8-branes either on top of the O8 or on circles inside $M_2$ as shown in figure \ref{fig:O8D8}. In analogy with \cite{seiberg-5d,brandhuber-oz,bah-passias-t}, one expects a configuration with $n_\mathrm{D8}$ D8-branes on the O8 to give rise to $E_{1+n_{\mathrm{D8}}}$ bulk gauge symmetry, and hence the putative dual CFTs would have exceptional flavor symmetry.

In the region near the O8 in our solution, the string coupling diverges, and the supergravity equations of motion are no longer physically relevant; they should be superseded by the complete string theory equations. While these are not known, at leading order in distance from the source our solution resembles the O8 solution in flat space, which presumably is a solution of string theory. This suggests that our solution should survive the onslaught of stringy corrections in the vicinity of the O8. Ideally one would be able to change to a different duality frame near the O8, and then patch this description with the supergravity solution which is valid almost everywhere.

A related general feature of our solutions is that, at the two-derivative level, they all possess a modulus that can be used to tune the solution to the reliable region of small curvature and weak string coupling. In other words, this modulus can be used to make the O8 region arbitrarily small.  However, the existence of this modulus also implies that their fate may be sensitive to higher derivative (stringy) corrections to the equations of motion as in the discussion of \cite{dine-seiberg}. Another way to think about this is that the stringy corrections are not invariant under shifts in the modulus.  This means that the stringy corrections generate a potential, and so we expect that our solutions survive at most for discrete values of the modulus.  We comment on this point in more detail in section \ref{sub:higher} below.

It is also natural to ask whether our solutions are stable. In contrast to more familiar supersymmetric solutions, non-supersymmetric AdS solutions are not a priori protected against instabilities. A conjecture has even been put forward \cite{ooguri-vafa-nonsusy-ads} that they are \emph{all} unstable, based on a certain non-perturbative decay channel mediated by brane bubbles. (For previous work on such bubbles see for example \cite{narayan-trivedi} and \cite[Sec.~4.1.2]{gaiotto-t}, which we will review below in footnote \ref{foot:gt}.)  Our solutions are simple enough that they can provide a simple playground to test such suspicions. This is especially important in view of our original motivation, namely to give evidence for the existence of unitary interacting CFTs in $d=7$.

We will argue that for our case the bubble instability of \cite{ooguri-vafa-nonsusy-ads} is not present. This is in part because the flux components along the internal volume, which are often present in an AdS solution, are not present in our solutions. Related to this, our solutions do not arise from any known near-horizon geometry, so one of the original motivations for the brane nucleation instability of \cite{ooguri-vafa-nonsusy-ads} seems inapplicable. 

One might wonder whether our solutions suffer from even more basic perturbative instabilities. One mode that we analyze in detail in section \ref{sub:stab} is uniform motion of the D8-branes in the internal space.  With a probe computation, we find that placing the D8-branes at generic positions in the internal manifold is in fact unstable.  Meanwhile the position of the D8-branes is stable if they are localized on top of the O8. Since the number of such D8-branes is bounded above $0 \leq n_{\mathrm{D8}}< 8$ this leads to a small list of candidate stable solutions.  A complete treatment of perturbative stability requires a Kaluza--Klein reduction, a challenging task which we leave for future work \cite{cordova-deluca-t-kk}.

We conclude in section \ref{sec:adsd} with a brief generalization of our section \ref{sec:o8-d8} analysis to other dimensions: this shows that there exist non-supersymmetric  AdS$_d$ solutions with only $F_0\neq 0$ and no other flux for other values of $d$ as well.  Parallel to our previous analysis, the simplest examples have as internal space a topological sphere $S^{10-d}$ with an SO(10-d) isometry, and an O8-plane at the fixed locus of an involution. 

Finally, appendix \ref{sec:other} contains a discussion of other candidate AdS$_{8}$ solutions which can either be excluded or are unphysical.

% section intro (end)
 
\section{O8--D8 Solutions} % (fold)
\label{sec:o8-d8}

We first look at IIA supergravity. In section \ref{sub:pre-eom} we specialize the equations of motion to our AdS$_8$ problem. We will immediately find that there are two options: $F_0\ne 0$, $F_2=0$ and $F_2\neq 0$, $F_2=0$. In this section we discuss explicit solutions for the first case. We examine the remaining case and IIB in appendix~\ref{sec:other}.

\subsection{IIA Equations of Motion} % (fold)
\label{sub:pre-eom}

The general IIA equations of motion are reviewed in appendix \ref{app:typeIIeom}. 

The most general ansatz preserving the isometries of AdS$_8$ is as follows. The metric can be written as 
\begin{equation}\label{eq:gen-metric}
	ds^2_{10}=e^{2W}ds^2_{{\rm AdS}_8} + ds^2_{M_2}~,
\end{equation}
where $M_{2}$ is the compactification two-manifold, and the warping function $W$ and dilaton are functions on $M_{2}$.  Throughout unless otherwise mentioned we fix conventions such that the cosmological constant $\Lambda$ is $-1.$ In IIA, the only possible fluxes consistent with our desired isometry are $F_0$ and $F_2$; the latter should be proportional to the volume form $\mathrm{vol}_2$ of the internal space $M_2$. Sometimes (for example in appendix \ref{app:typeIIeom}) we also refer to their duals defined by (\ref{eq: typeIIeom:5}). 

From the flux equations of motion, we immediately notice some strong constraints. Throughout our solution, the NS-NS three-form $H$ must vanish.  This means that, away from brane sources, we have $dF_p=0$ (see (\ref{eq: typeIIeom:4})).  Comparison with  \eqref{eq: typeIIeom:3} then implies that
\begin{equation}\label{eq:F0F2}
	F_0 \wedge *F_2 = 0 ~.
\end{equation}
Thus, in IIA our analysis will split in the two cases $F_0\neq0$, $F_2=0$ and $F_2 \neq 0$, $F_0=0$. 

In order to get concrete results, we will mostly consider the cohomogeneity-one ansatz, defined by taking the metric to be
\begin{equation}\label{eq:coh1}
	ds^2_{10}=e^{2W}ds^2_{{\rm AdS}_8} + e^{-2Q}(dz^2 + e^{2\lambda} d\theta^2)\,,
\end{equation}
where $\theta\sim \theta +2\pi$ is periodic and now $W$, $Q$ and $\lambda$ as well as the fluxes and dilaton only depend on $z$. We can fix the radial ($z$) reparameterization gauge freedom for example by fixing $Q$ in terms of other functions.\footnote{We can assume $e^\lambda >0$, even if only its square appears in the metric. In a solution where $e^\lambda$ changes sign, it goes through zero; at such a point the $S^1$ shrinks and the manifold ends (see below).}                                                                                                      

% subsection eom (end)

\subsection{Reduction to Ordinary Differential Equations} % (fold)
\label{sub:eom}

In this section we will solve the condition (\ref{eq:F0F2}) by taking $F_2=0$. (The other possibility of $F_2\neq 0$, $F_0=0$ is discussed in section \ref{sub:F00}.)

We specialize the general type II equations of motion \eqref{eq:typeIIeom} to the ansatz (\ref{eq:coh1}).  To begin we work away from brane sources, and thus neglect localized $\delta$ terms. We will introduce sources in section \ref{sub:dw} below. We find the following system of ODEs (below prime indicates derivative with respect to $z$):
\begin{subequations}\label{eq:zcoh1}
	\begin{align}
		&-4 e^{-2(Q+W)} = 2(\phi')^2 + \lambda'(\lambda- Q - 2 \phi)' +4 W' (9 W' +2\lambda - 4 \phi)'  +(\lambda-2\phi +  8W-Q)'' \label{eqz:1}\\
		&4 e^{-2(Q+W)}= e^{2(\phi-Q)}F_0^2 -4 W'\left(\lambda -2\phi + 8W\right)'-4 W'' \label{eqz:2}\\
		&0=\frac{1}{4}e^{2(\phi-Q)}F_0^2 - (\lambda')^2 - 8 (W')^2 + Q'(-8 W + \lambda +2 \phi)' - (\lambda-2\phi + 8W-Q)''
		\label{eqz:3}\\
		&0=\frac14 e^{2(\phi-Q)}F_0^2 - (\lambda')^2 - 8 W' \lambda' + 2 \lambda' \phi' +Q'(8W+ \lambda -2 \phi)' + (Q-\lambda)''\label{eqz:4}
	\end{align}
\end{subequations}
The coordinate $z$ never appears explicitly in the equations: in other words, the system is autonomous. 

As usual in general relativity and in theories with gauge redundancies, from the system \eqref{eq:zcoh1} we can extract a first-order linear combination:\footnote{One can try to generate further first-order equations by taking a first derivative of \eqref{eq:first} and subtracting the second derivatives using (\ref{eq:zcoh1}). However, this putative new equation is in fact proportional to (\ref{eq:first}). }
\begin{equation}\label{eq:first}
	-2 e^{-2(Q+W)} = \frac18 e^{2(\phi-Q)}F_0^2 + 2 W'(7W-2Q+2 \lambda)' + \phi'(Q-8W- \lambda)'+(\phi')^2\,.
\end{equation}
We can trade an equation appearing in the combination, say \eqref{eqz:1}, for (\ref{eq:first}). Moreover, another equation, say \eqref{eqz:4}, is a linear combination of $\partial_z$(\ref{eq:first}), (\ref{eq:first}), \eqref{eqz:2} and \eqref{eqz:3}.\footnote{Actually this is true just if $	\lambda' \neq Q'$. But if we took $\lambda' = Q'$, then \eqref{eqz:4} would set $F_0 = 0$ and there are no solutions.} This leaves us with a system of three equations: (\ref{eqz:2}), (\ref{eqz:3}), (\ref{eq:first}). 

Observe also that $\lambda$ never appears underived in this system. Therefore, given any solution, we can obtain another by shifting $\lambda$ by a constant. Below we will consider smooth points where the circle parameterized by $\theta$ collapses.  In this case smoothness of the solution fixes this freedom.

We can achieve some further simplification by fixing the radial reparametrization gauge freedom in (\ref{eq:coh1}) with the choice 
\begin{equation}\label{eq:Q=W}
	Q= W \,,
\end{equation}
so that the metric now reads
\begin{equation}\label{eq:gf}
ds^2_{10} = e^{2W}ds^2_{\text{AdS}_{8}}+ e^{-2W}(dz^2 + e^{2\lambda}d\theta^2) \,.
\end{equation}
This gauge is often useful in other contexts, including for AdS$_7$ solutions \cite{cremonesi-t} and for black hole solutions in general relativity. With the further definition 
\begin{equation}\label{eq:alpha}
	\alpha= e^{\lambda - 2 \phi + 8 W} ~,
\end{equation}
we obtain the system 
\begin{subequations}\label{eqz4}
	\begin{align}
		&0=2e^{-4W} + \frac{F_0^2}{8}e^{2\phi - 2W} - \frac{\alpha'}\alpha(\phi-4W)'-\left((\phi')^2 -9\phi'W' + 22(W')^2\right) \,,   \label{eqz4:1}\\
		%&0=\alpha e^{-4W} - 2\alpha'(2\phi-9W)'-\alpha''-\alpha\left(4(\phi')^2 -36\phi'W' + 88(W')^2-2W''\right) 
		%&e^{4\phi} = - \frac{16\left(\alpha'' - 9 \left(\alpha W'\right)'\right)^2}{9 F_0^{\,4}\,\alpha \alpha''}\\
&e^{2\phi} = 4 e^{-2W}\frac{\left(\alpha+e^{4W}(\alpha W')'\right)}{F_0^2 \alpha}  \,,
		\label{eqz4:2}\\
		&e^{4W} = -9\frac{\alpha}{\alpha''}\,.\label{eqz4:3}
	\end{align}
\end{subequations}
Up to factors, the last equation determines the warping function (the coefficient of $ds^2_\mathrm{AdS}$) in the same way obtained for AdS$_7$ solutions in \cite[(2.27)]{cremonesi-t}.  We also note that, since $\alpha$ is positive, equation \eqref{eqz4:3} implies that $\alpha''$ is negative definite.  In particular this means that the geometry $M_{2}$ cannot be periodic.

Another important feature of the system (\ref{eqz4}), is that it is invariant under the rescaling
\begin{equation}\label{eq:resc}
	W \to W + c \, ,\qquad \phi \to \phi - c \, ,\qquad
	\lambda \to \lambda + 2c \, ,\qquad z \to e^{2c} z~.
\end{equation}
This can equivalently be thought of as
\begin{equation}\label{eq:resc-10}
	ds^2_{10} \to e^{2c} ds^2_{10} \, ,\qquad e^{\phi}\to e^{-c} e^\phi\,.
\end{equation}
Given any solution, (\ref{eq:resc}) can be used to generate another solution with smaller curvature and smaller string coupling $e^\phi$, without changing $F_0$. Thus we can get parametrically good perturbative control over any solution.
(Taking into account higher order corrections, one expects this modulus to be lifted; see section \ref{sub:higher}.)

Unfortunately we have not found analytic solutions to this system of ODEs, but in the following we will see that one can straightforwardly find numerical solutions. Before proceeding we record one further way of writing the equations of motion:
\begin{equation}
	\begin{split}
 \Delta_f W &= -\frac{1}{4}e^{2\phi}F_0^2+e^{-2W}\,,\\
 \Delta_f \phi &= 5\Delta_f W -5e^{-2W}= -\frac{5}{4}e^{ 2\phi}F_0^2\,,
	\end{split}
\end{equation}
with $\Delta_f $ a modified Laplacian
\begin{equation}
	\Delta_f y \equiv -e^{-f}\nabla^\alpha \left(e^{f}\nabla_\alpha y\right)\;,\qquad f \equiv 8W - 2\phi\;.
\end{equation}
The covariant derivatives are computed with respect to the purely two-dimensional metric $ds^2_{{M}_2}$.

% subsection eom (end)

\subsection{Domain Wall Conditions} % (fold)
\label{sub:dw}

We now consider what happens near sources.  Since only $F_{0}\neq 0,$ we restrict our attention to D8-branes and O8-planes. Fortunately the singularities that these objects induce in the fields are relatively mild and we can treat them using simple distributional derivatives.

A first subtlety is that the process of elimination we performed in section \ref{sub:eom} does not quite work in the same way; this is basically because $F_0$ is not constant, and its derivative generates additional $\delta$s. An alternative presentation of the system is then
\begin{subequations}\label{eq:eom-delta}
\begin{align}
 &\frac18F_0^2 e^{2 (-W+\phi )}=\frac{\alpha'}\alpha  \left(\phi-4 W\right)'+  \left(-9 W' \phi'+22 \left(W'\right)^2+\left(\phi '\right)^2\right)-2 e^{-4W}\;, \label{eq: eqz2So:1} \\
 &\frac14 e^{-2 W+\phi } \left(F_0^2 e^{\phi }+e^W \kappa ^2 \tau  \delta \left(z-z_0\right)\right)=W''+W' \frac{\alpha '}\alpha + e^{-4 W}\;, \label{eq: eqz2So:2}\\
 %&0=e^{2 W+\phi } \left(F_0^2 e^{\phi }-\kappa ^2 \tau  \delta \left(z-z_0\right)\right)-\frac{4 e^{4 W} \left(\alpha  W'\right)'}{\alpha }-4 \\
 &e^{4W} = -9\frac{\alpha}{\alpha''}\;, \label{eq: eqz2So:3}\\
 %&\left(\phi-5 W\right)''+\frac{\alpha'}\alpha \left(\phi-5 W\right)'+\frac{\alpha''}\alpha = \frac12 e^{-4W} \;. \label{eq: eqz2So:4}
&-\frac54 e^{-2W+\phi}\left(F_0^2 e^{\phi} + e^{W} \kappa^2 \tau \delta\left(z-z_0\right)\right) = \phi''+\phi'\frac{\alpha'}{\alpha}  \;. \label{eq: eqz2So:4}
\end{align}
\end{subequations}
Finally, the Bianchi identity for $F_0$ tells us that it jumps at brane sources according to
\begin{equation}\label{eq: eomF0sources}
	d F_0 = -  \kappa^2 \tau \delta(z-z_0)\;.
\end{equation}

The presence of localized sources in \eqref{eq:eom-delta} makes it clear that the functions defining the solution are no longer smooth.  Instead, we assume that the variables are continuous but not differentiable: in a distributional (or weak) sense, their first derivatives are discontinuous, and their second derivatives have some $\delta$ terms. For example, the function $|z-z_0|$ has a weak first derivative $\partial_z |z-z_0| = \mathrm{sign}(z-z_0)$, and a weak second derivative $\partial_z^2 |z-z_0| = \partial_z \mathrm{sign}(z-z_0) = 2 \delta(z-z_0)$.

Let us examine the behavior of the variables at a source locus $z=z_0$ (where the first derivatives are discontinuous). The discontinuity in the first derivative can be determined by integrating (\ref{eq:eom-delta}) on an infinitesimal interval around $z_0$. We obtain:
\begin{subequations}\label{eq:dw}
\begin{align}
	&e^{W-\phi}\Delta W' = \frac{1}{4}\kappa^2\tau \, ,\qquad 
	e^{W-\phi}\Delta\phi' = \frac{5}{4}\kappa^2\tau  \, ,\qquad\Delta \alpha' = 0  \, ,\label{eq:jumps}\\
	&	F_0 =-e^{W-\phi} \left(\phi'- W' + \frac{\alpha'}\alpha\right) \,.   \label{eq:stop}
\end{align}	
\end{subequations}
All quantities which are not under the variation sign $\Delta$ are to be understood as evaluated on the left of the object, i.e.\ for $z\to (z_0)^-$. (We have used $\Delta (F_0^2)= 2 F_0 \Delta F_0 + (\Delta F_0)^2$.) Meanwhile, using (\ref{eq:kappa}), (\ref{eq:tau's}) and (\ref{eq: eomF0sources}), in string units $l_s = 1$,
\begin{equation}\label{eq:k2t}
	-\kappa^2 \tau =  - \frac1{2\pi}(n_{\mathrm{D8}} - 8 n_{\mathrm{O8}} ) = \Delta F_0 =  \frac{\Delta n_0}{2\pi}\,,
\end{equation}
where $n_\mathrm{O8}\in \{ 0, 1\}$. 
%From $\Delta(F_0^2)$ we can extract the jump of $F_0$: $\Delta F_0 = -F_0 \pm \sqrt{F_0^{\,2} + \Delta(F_0^{\,2})}$. This expression has to agree with the jump of $F_0$ obtained by integrating its equation of motion \eqref{eq: eomF0sources} across the sources, namely $\Delta F_0 = - \kappa^2 \tau$, which gives  
%\begin{equation}\label{eq: dn0}
%	\Delta n_0 \equiv 2\pi \Delta F_0 = -(n_{D8} - 8 n_{O8} )\,.
%\end{equation}

For example, we can apply this to an O8-plane (possibly with $n_\mathrm{D8}<8$ D8-branes on top). By definition of O8, the solution has an involution relating the left and right sides of the O8-plane. Then we have for example $\Delta F_0 = -2 F_0$, where $F_0$ is the value of the Romans mass on the left side.  (\ref{eq:dw}) then become 
\begin{equation}\label{eq:dw-O8}
\begin{split}
	-2 e^{W-\phi}\phi'|_\mathrm{O8} = -10 e^{W-\phi}W'|_\mathrm{O8} =   \frac54 \kappa^2\tau  \, ,\qquad
	\alpha'|_\mathrm{O8} = 0  \, ,
\end{split}	
\end{equation}
where again all quantities are evaluated on the left side $z<z_\mathrm{O8}$ of the O8; $\kappa^2\tau$ is given by (\ref{eq:k2t}) (with $n_\mathrm{O8}=1$). Using (\ref{eq:dw-O8}) in (\ref{eq: eqz2So:1}) we get
\begin{equation}\label{eq:no-fin-O8}
	\Lambda e^{-4W}|_\mathrm{O8}=0~,
\end{equation}
where in the above we have restored the cosmological constant $\Lambda$.

To verify that these discontinuities indeed describe an O8-plane we should check that locally our metric behaves like that of an O8-plane in flat space: 
\begin{equation}\label{eq:O8loc}
	ds^2_{10} \sim H^{-1/2}(-dx_0^2 + dx_1^2 +\ldots dx_8^2)+ H^{1/2} dx_9^2\,,
\end{equation} 
where $H= a + b |x_9|$ for some $a\ge 0$ and $b= \frac{g_s}{4\pi} (8-n_{\mathrm{D}8})$. Comparing with (\ref{eq:coh1}) we see $H=e^{-4W}$; so we deduce from (\ref{eq:no-fin-O8}) that we can only have O8-planes for which $a=0$. In this case, the string coupling $e^\phi\sim g_s H^{-5/4}$ diverges.  (By contrast in flat space $\Lambda$ vanishes and \eqref{eq:no-fin-O8} does not constrain the warp factor.)

This result is not entirely surprising: in most AdS solutions in other dimensions, the O8-planes that appear have a diverging dilaton (see for example \cite{brandhuber-oz,bah-passias-t}).

In a region where the dilaton diverges, however, the logic that took us to (\ref{eq:dw}) should be reexamined. At a formal level, we cannot really use the weak second derivative $\partial_z^2 |z-z_0| = 2 \delta(z-z_0)$, since the functions diverge rather than just having an angular point. Various formal manipulations can be attempted; one is for example to change variables to ones that still have an angular point, such as $H_1=e^{-4W}$ and $H_2=e^{-\frac45 \phi}$. This takes us back to (\ref{eq:dw}); other changes of variables however might take us to impose (\ref{eq:dw}) even at subleading orders in $|z-z_0|$. Most importantly, however, while in this paper we use supergravity as a tool, we are ultimately interested in finding solutions that are valid in fully-fledged string theory. In the region where the dilaton diverges, the supergravity equations of motion are no longer valid, and strictly speaking we cannot use the logic leading to (\ref{eq:dw}) at all. (We can use (\ref{eq:resc}) to make this strongly-coupled region as small as we like, but we can never eliminate it completely.) In spite of this, we will still use (\ref{eq:dw}) as a domain-wall condition even if the dilaton diverges, given that it reproduces the same leading-order behavior as an O8 in flat space. We will return to this point in section \ref{sub:higher}.

% subsection dw (end)

\subsection{Perturbative Solutions} % (fold)
\label{sub:pert}

We will now study the equations of motion in a power-series approach. 

First we look for a solution for which the circle with coordinate $\theta$ shrinks at some point, so that the space is regular (non-singular) around it. Without loss of generality, we can take this point to be at $z=0$. Regularity then means that
\begin{equation}\label{eq:reg}
	e^\lambda = z + O(z^2) \, ,\qquad W= W_0 + O(z^2) \, ,\qquad \phi = \phi_0 + O(z^2)\,.
\end{equation}
The behavior of $e^\lambda$ is so that the internal metric in (\ref{eq:coh1}) behaves as $ds^2_{10}\sim dz^2 + z^2 d \theta^2+ \ldots$, which is the $\mathbb{R}^2$ metric when the periodicity $\Delta \theta= 2\pi$. (In particular this fixes the freedom in shifts of $\lambda$ mentioned below \eqref{eq:first}) The absence of linear terms in $W$ and $\phi$ is so that they are at least $C^2$, since $z$ is a radial coordinate. Solving the equation of motion \eqref{eqz4} assuming (\ref{eq:reg}) leads to
\begin{subequations}\label{eq: pertRegular}
\begin{align}
	W = & \frac{\log \left(c_1\right)}{2}+\frac{z^2 \left(c_0^2 c_1 F_0^2-4\right)}{16 c_1^2}+\frac{z^4 \left(c_0^2 c_1 F_0^2 \left(c_0^2 c_1 F_0^2+6\right)-20\right)}{128 c_1^4}+\nonumber\\
	+&\frac{z^6 \left(c_0^2 c_1 F_0^2 \left(30 c_0^4 c_1^2 F_0^4+99 c_0^2 c_1 F_0^2+1142\right)-3072\right)}{23040 c_1^6}+O\left(z^7\right)\,, \\
 \phi = &\log \left(c_0\right)+\frac{5 c_0^2 F_0^2 z^2}{16 c_1}+\frac{5 c_0^2 F_0^2 z^4 \left(c_0^2 c_1 F_0^2+4\right)}{128 c_1^3}+\nonumber\\
  + &\frac{c_0^2 F_0^2 z^6 \left(30 c_0^4 c_1^2 F_0^4+99 c_0^2 c_1 F_0^2+598\right)}{4608 c_1^5}+O\left(z^7\right)\,, \\
 \alpha = &\frac{c_1^4 z}{c_0^2}-\frac{3 c_1^2 z^3}{2 c_0^2}+\frac{9}{80} z^5 \left(c_1 F_0^2+\frac{2}{c_0^2}\right)+O\left(z^7\right)~.
\end{align}
\end{subequations}
The results depend on two real parameters $c_0$, $c_1$. 

In fact, imposing that $\alpha$ has the form in (\ref{eq:reg}) already implies the other two conditions there, namely that $W'$ and $\phi'$ vanish at $z=0$. Our system (\ref{eqz4}) consists of two second-order equations and one first-order equations; so at a generic point one expects five parameters to determine the initial conditions. One might then expect that imposing two conditions should result in a local solution with three free parameters.  To see why we instead only have two free parameters in (\ref{eq: pertRegular}), notice that (\ref{eqz4:2}) near $z=0$ has the form $\alpha (W''+ ...) = W' \alpha'$, and so setting $\alpha = 0$ implies $W'=0$ automatically; this fixes one extra parameter. This phenomenon can also be understood in the framework of quasi-linear systems of ODEs, namely systems of the form $M(q) q' = v(q)$, where $q$ is a vector of variables, and $M$ and $v$ are a matrix and vector which can be taken to depend on $q$ nonlinearly. Our system can be cast in this form in terms of the three variables $\alpha$, $\phi$, $W$ and their first derivatives; it becomes a point-dependent vector field on the five-dimensional space ${\mathcal M}_5$ defined by the first-order equation (\ref{eqz4:1}). At a generic point on ${\mathcal M}_5$, the matrix $M(q)$ is invertible and $q'= M(q)^{-1} v(q)$ is determined; at special loci where $M(q)$ is non-invertible, however, the system will have no solution unless we impose $v(q)$ to be in the image of $M(q)$, and this fixes more parameters than one originally intended.

Another local behavior that will be interesting for us is that around an O8 (with $n_\mathrm{D8}$ D8-branes on top). From the discussion around (\ref{eq:O8loc}) we have the local behavior
\begin{equation}\label{eq:O8local}
	e^\lambda \sim t^{-1/2} \, ,\qquad e^W \sim t^{-1/4} \, ,\qquad e^\phi \sim t^{-5/4}
	\, ,\qquad t \equiv |z-z_0|\,.
\end{equation}
Moreover, we also note from (\ref{eq:alpha}) that $\alpha$ goes to a constant. 

In a strong coupling region the supergravity equations of motion are not physically relevant, as we discussed at the end of the previous subsection. In spite of this, we can try as an exercise to formally solve the supergravity equations of motion in the neighborhood of the O8. Identifying the subleading behavior in (\ref{eq:O8local}) is not immediate: one has to decide for example if the expansion parameter is $t$ or some fractional power like $t^{1/4}$. (A similar problem presented itself for the O6 in \cite[(5.8)]{rota-t}.) After some experimentation and some help from numerical results, we obtain
\begin{subequations}\label{eq: pertO8}
		\begin{align}
		e^{4W}&= \frac{a_2}{t}+a_1+\frac{5 a_1^2 t}{3 a_2}+\left(\frac{4 a_1^3}{3 a_2^2}-\frac{41}{24}\right)t^2 \\
	&+\left(\frac{71 a_1^4}{45 a_2^3}-\frac{43 a_1}{60 a_2}\right) t^3+\left(\frac{52 a_1^5}{45
   a_2^4}-\frac{517 a_1^2}{360 a_2^2}\right) t^4+O\left(t^5\right)\,, \nonumber\\
	 e^{4\phi}F_0^4 & = \frac{a_2}{t^5}+\frac{a_1}{t^4}+\frac{19 a_1^2}{3 a_2 t^3}+\frac1{t^2}\left(\frac{6a_1^3}{a_2^2}-\frac{125}{24}\right) \\
	&+\frac{256 a_1^4-41 a_1 a_2^2}{12 a_2^3 
   t}+\frac{6968 a_1^5-10499 a_1^2 a_2^2}{360 a_2^4 } +O\left(t^1\right)\,, \nonumber\\
			\alpha &= a_0-\frac{3 a_0 t^3}{2 a_2}+\frac{3 a_0 a_1 t^4}{4 a_2^2}+\frac{3 a_0 a_1^2 t^5}{10   a_2^3}-\frac{a_0 \left(24 a_1^3+5 a_2^2\right) t^6}{80 a_2^4}+O\left(t^8\right)~, \label{alphao8eq}
		\end{align}
\end{subequations}
where $a_0$, $a_1$, $a_2$ are three real parameters. The O8 domain-wall conditions (\ref{eq:dw-O8}) are automatically satisfied by this solution; remarkably, even the correct coefficient $-\kappa^2 \tau= \Delta F_0 = -2 F_0$ is reproduced. This means that the bulk supergravity equations already know about the correct O8 tension, even without imposing supersymmetry.

% subsection pert (end) 

\subsection{Numerics} % (fold)
\label{sub:num}

We can now use the perturbative solutions we found as a seed for a numerical study. 

For example, we can start from a regular point. The usual technique is to evaluate the perturbative solution (\ref{eq: pertRegular}) at a small value of $z$, where it is very reliable, and use it as an initial condition for a numerical evolution using the system (\ref{eqz4}).\footnote{\label{foot:pert}In fact, it is also possible to push (\ref{eq: pertRegular}) to very high order, obtaining results which are virtually indistinguishable from the numerical solutions.} 

For each value of the initial conditions, there are actually two possible solutions: this is due to the first-order equation (\ref{eqz:1}), which is quadratic in the first derivatives. The result of the numerical evolution always results in a singularity for both solutions. But one of these singularities is exactly (\ref{eq:O8local}),\footnote{\label{foot:sing}The other type of solution has a singularity for which $e^W\sim t^{2/11}$, $e^\phi\sim t^{7/11}$, $e^\lambda\sim t^{9/11}$; we cannot match this to any IIA object, and we thus conclude that it is unphysical and do not consider it further.} the back-reaction of an O8 with diverging dilaton, possibly with $n_\mathrm{D8}<8$ D8-branes on top.

To be more precise, the solution one gets this way is not physical for any choice of the initial parameters: a fine tuning is required to reproduce the correct local behavior of $\alpha$ in \eqref{alphao8eq} (specifically we must reproduce $\alpha'\to 0$ at the singularity).\footnote{Note from the local behavior \eqref{alphao8eq}, that near an O8  both $\alpha'\to 0$ and $\alpha''\to 0.$  However, the equation of motion \eqref{eq: eqz2So:3} shows that as soon as $W$ diverges $\alpha''$ automatically vanishes.  Thus only a one parameter tuning is necessary to achieve the correct local behavior of the O8.  } The resulting solutions depend on a single real parameter which can be identified with the modulus discussed in (\ref{eq:resc}) and can be used to make $e^\phi$ and the curvature of this solution arbitrarily small.

\begin{figure}[ht]
	\centering
		\includegraphics[width=8cm]{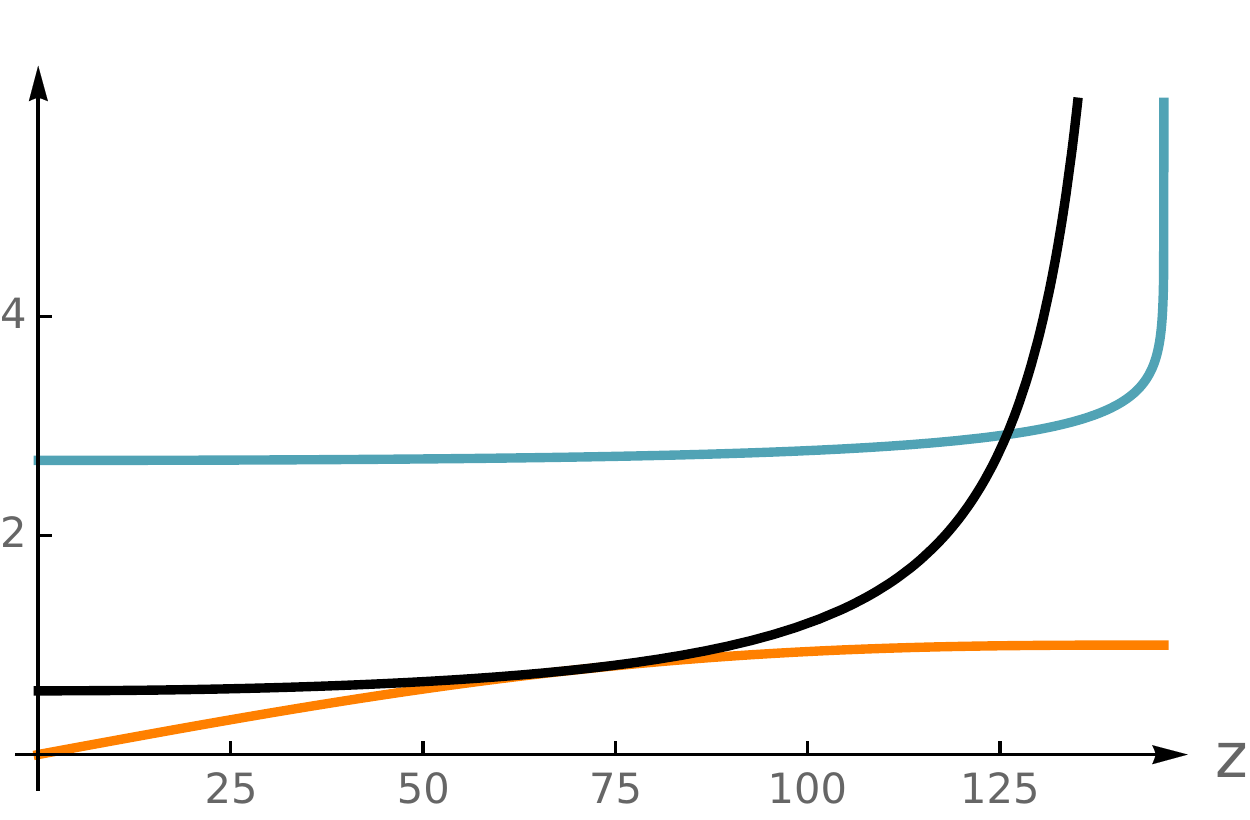}
	\caption{\small A numerical solution with $e^\phi$ (black), $W$ (turquoise), $\alpha/\alpha_{\mathrm{O8}}$ (orange), with $F_0=-\frac{2}{2\pi}$, $n_\mathrm{D8}=4$. At the left endpoint, the solution is regular; at the right endpoint, it behaves as an O8 with diverging dilaton. (One should imagine a mirror copy of the solution to the right of the O8; the two halves are identified by the orientifold involution.)}
	\label{fig:reg-O8}
\end{figure}

On the other hand, the rest of (\ref{eq:dw-O8}) works automatically, given the Bianchi identity, which from (\ref{eq:k2t}) in this case reads $-2n_0 = \Delta n_0= - (n_\mathrm{D8}-8)$. This is a consequence of our remark below (\ref{eq: pertO8}), where we noticed that any solution with the correct local behavior (\ref{eq:O8local}) already reproduces the correct O8 tension.

Even though we are only showing in figure \ref{fig:reg-O8} the solution on the left of an O8, one should imagine a mirror copy of it to the right of the O8. The two halves are identified by the orientifold involution.

We can also try to insert some D8-branes which are not on top of the O8. We again use (\ref{eq: pertRegular}) as a starting point. We can place D8-branes at a locus where (\ref{eq:stop}) is satisfied.  For convenience we rewrite this condition as
%, but we now stop the evolution at a point $z_{\mathrm{D8}}$ where (\ref{eq:stop}) is satisfied. %This is the locus where the D8-branes will be placed. Notice that the value of $z_\mathrm{D8}$ does not depend on the number of D8-branes that we want to place there; this is similar to what found for AdS$_7$ solutions, where the positions of the D8-branes depend on their D6-charge and not on their number.
%For later convenience let us rewrite (\ref{eq:stop}) as 
\begin{equation}\label{eq:stop2}
n_0 = - 2 \pi \frac{e^{W - \phi}}{\alpha}\Big{(}\alpha (\phi' - W') + \alpha'\Big{)}\Big{\rvert_{z = z_{\text{D8}}}}
\end{equation}
We can stop the evolution at the point $z=z_\mathrm{D8}$ where the above is satisfied, and we place $n_{\text{D8}}$ D8-branes there. We then start the evolution again from this point, with new initial data obtained by applying (\ref{eq:jumps}). The solution we obtain again leads to a diverging-dilaton O8. As in the case without D8-branes we have to fine tune our initial conditions at $z=0$ so that $\alpha'\rvert_{\text{O8}} = 0$ is satisfied. 

From this procedure we can see some restrictions on the number of D8-branes in our geometry.  Specifically, we can show that the initial $n_{0}$ in equation  \eqref{eq:stop2} cannot be positive.  To verify this, first note that a combination of the equations of motion can be rewritten as
\begin{equation}
\partial_{z} \big( \alpha(\phi'-W')\big) = \alpha e^{-4W}\left(e^{2\phi-2W} F_0^2+1\right)\geq 0\,.
\end{equation}
The function $\alpha(\phi'-W')$ vanishes at a regular point, and hence by the above is everywhere non-negative.  This implies that \eqref{eq:stop2} can only be satisfied for positive $n_{0}$ if $\alpha'\rvert_{z = z_{\text{D8}}} < 0.$  However, as remarked below equation \eqref{eqz4}, the equations of motion imply that $\alpha'' \leq 0.$  Thus if $\alpha'$ is negative, it will continue to decrease, and can never reach its required value of zero on the O8.  

Let us combine this argument with the Bianchi equation which relates $n_0$ to the number $n_{\text{D8}}$ of D8-branes away from the O8, and to the number $\tilde{n}_{\text{D8}}$ of (half-)D8-branes on top of the O8:
\begin{equation}\label{eq:n0-concl}
n_0 = n_{\text{D8}}+\frac{\tilde{n}_{\text{D8}}}{2}-4 < 0 \,.
\end{equation}
We conclude that there is an upper bound on the total number of D8-branes we can place in our solution.

\begin{figure}[ht]
	\centering
		\includegraphics[width=8cm]{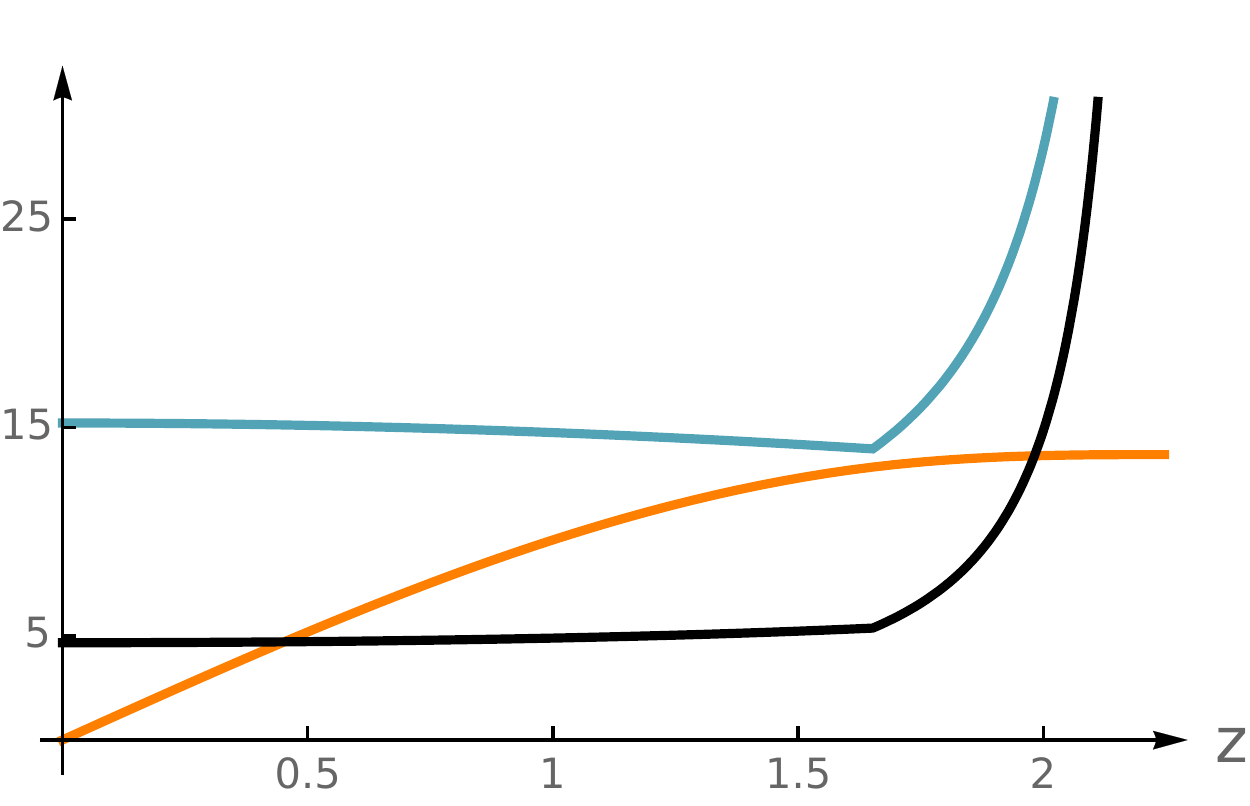}
	\caption{\small A numerical solution with a single stack of three D8-branes, to the left of a diverging-dilaton O8.  The values of the Romans mass are $F_0=\frac{n_0}{2\pi}$, with $n_0=(-1,-4)$. To illustrate the kink we have plotted $\alpha$ (orange), $e^{4W}$ (turquoise) and $e^{\phi}$ (black).}
	\label{fig:reg-D8-O8}
\end{figure}

In principle, this process can be repeated to obtain solutions with several stacks of D8-branes. In figure \ref{fig:reg-D8-O8} we show an example with a single stack of D8-branes and a diverging-dilaton O8 to its right. The D8 stack manifests itself as the angular point in the functions, where they are continuous but their derivatives change. 

One might also try to consider anti-D8-branes. An anti-D8 has the opposite charge of a D8 but the same tension and thus are not just obtained by considering a negative $n_\mathrm{D8}$ above. Taking this into account one sees that the left-hand side of (\ref{eq:stop2}) changes sign; one now concludes that $n_0$ on the left is positive, rather than negative as in (\ref{eq:n0-concl}). This would require a sufficient number of half-D8-branes on top of the O8 so that the total O8-D8 system has positive tension. This is in contradiction with what we can have in the solution, and we conclude that  anti-D8s are forbidden.

% subsection num (end)

\subsection{Higher-Derivative Corrections} % (fold)
\label{sub:higher}

As mentioned around equation \eqref{eq:resc-10}, a general feature of all our solutions is that they come in one-parameter families obtained by acting on any solution with the transformation:
\begin{equation}\label{eq:resc-repeat}
	ds^2_{10} \to e^{2c} ds^2_{10} \, ,\qquad e^{\phi}\to e^{-c} e^\phi\,.
\end{equation}
Scaling to large $c$, the solutions become weakly curved and have small string coupling $e^{\phi}$.  Although the rescaling modulus \eqref{eq:resc-repeat} is classically a flat direction, it is natural to expect that quantum corrections to the IIA string theory effective action will lift this mode.  This is especially true given that our solution is non-supersymmetric and hence there is no symmetry reason for a flat modulus to persist.  

This issue is related to the fact that near the O8 the string coupling diverges. As we mentioned at the end of section \ref{sub:dw}, in this region the supergravity equations of motion are superseded by the unknown equations of motion of string theory. While we have no access to those equations, our solution resembles at leading order in $|z-z_0|$ the O8 solution in flat space, which should exist in string theory, given its fundamental definition in terms of open strings. This gives us good hope that our solution also exists in full string theory. However, the equations of motion of full string theory are not invariant under (\ref{eq:resc-repeat}); presumably, then, the solution should only be valid for one particular value of $c$.

In spite of these difficulties, we could try to proceed as follows. One approach to these corrections is to evaluate the higher derivative terms on a given family of solutions and thereby view them as a generating an effective potential for the mode $c$.  This method is reliable in the regime of large $c$ where the corrections are small, and leads to a qualitative picture analogous to that discussed in \cite{dine-seiberg}.  For instance, the leading order corrections to the IIA effective action are tree level in the string coupling and begin with $R^{4}$ (see e.g.\ \cite{policastro-tsimpis, liu-minasian} for a recent summary).\footnote{There are also higher derivative corrections to the brane worldvolume actions that we neglect in our qualitative discussion below.}  Schematically
\begin{equation}\label{eq:R4}
S^{\rm tree}_{R^{4}}\sim \int d^{10}x\sqrt{g_{10}}~e^{-2\phi}\left(t_{8}t_{8}+\frac{1}{8}\epsilon_{10}\epsilon_{10}\right)R^{4}+\cdots~,
\end{equation}
where above the terms $t_{8}t_{8}$ and $\epsilon_{10}\epsilon_{10}$ indicate particular index contractions of the Riemann tensor, and $\cdots$ include for instance terms with derivatives of the dilaton.  The parametric dependence of (\ref{eq:R4}) on the modulus $c$ is easy to fix based on scaling and is simply $e^{4c}$.  Similarly, at next order there are one-loop $R^{4}$ as well as tree level $R^{5}$ terms, which scale as $e^{2c}$.

To deduce the effective potential we rewrite these corrections in the eight-dimensional effective action in the form
\begin{equation}
S_{\rm effective}\sim \int d^{8}x\sqrt{g_{8}} \ (R-V(c))~,
\end{equation}
and hence from our qualitative discussion above we have (restoring the cosmological constant)
\begin{equation}
V(c)=\Lambda +A e^{-4c}+B e^{-6c}+\cdots\,.
\end{equation}

There are several essential challenges to making this approach quantitatively reliable even in the regime of large $c$.  The first is a question of practice: although much work has been done on higher derivative corrections to supergravity, the complete form of even the leading order corrections including all relevant terms is not explicitly known.  A second challenge is one of principle.  The coefficients $A$ and $B$ in the potential above should be determined by evaluating the various curvatures on our compactification manifold.  However, all our solutions have O8 sources near which the curvatures diverge. In principle this means that the higher order terms in the effective action become relevant. As we discussed above, the fact that the O8 is an exact solution of string theory suggests that this is not a fundamental challenge; but it does make it difficult to treat the higher derivative corrections systematically.

% subsection higher (end)

\subsection{Stability} % (fold)
\label{sub:stab}
  
As we anticipated in the introduction, it is natural to wonder about the stability of our solutions. We will first consider the solutions with the O8-plane only, and then consider solutions with D8-branes at the end of the subsection.

In general there are two possible types of instabilities: perturbative and non-pertur\-ba\-tive. The first can be assessed with a Kaluza--Klein reduction around the solution, which we will present elsewhere \cite{cordova-deluca-t-kk}. The second occurs when a tunneling event  at a point in spacetime takes the fields to a different vacuum; this generates a bubble which can then expand and reach the boundary of AdS in finite time. 

A first type of bubble that one can consider is a D-brane domain wall. For an AdS$_d\times M_{10-d}$ compactification, this would be a D$p$-brane wrapping an $\mathbb{R}\times S^{d-2}\subset$ AdS$_d$ (with the $\mathbb{R}$ direction being time) and a $(p-d+2)$-submanifold $\subset M_{10-d}$. Given that the RR flux jumps across a brane, the vacua inside and outside will not be the same: the brane represents a domain wall connecting two different vacua. Assuming such a brane is created by a quantum effect, we can wonder whether the $S^{d-2}$ will expand or collapse. The D-brane action contains a gravitational DBI term, which will make the brane collapse, and a coupling to the RR fields, which in general will want to make it expand (much like an electron-positron pair in an electric field, in the Schwinger effect). In supersymmetric compactifications, these two can exactly cancel each other, in which case the brane represents a BPS domain wall. In non-supersymmetric compactifications, one of the two terms will dominate. A natural extension of the weak-gravity conjecture \cite{arkanihamed-motl-nicolis-vafa} suggests that there is always a brane for which the gravitational term is weaker, which will make it expand \cite{ooguri-vafa-nonsusy-ads}.\footnote{\label{foot:gt}For an illustration of this mechanism, see for example the non-supersymmetric AdS$_4$ vacua in \cite[Sec.~4.1]{gaiotto-t}. For those vacua, the computation in section 4.1.2 there shows that D2-branes wrapping $\mathbb{R}\times S^2 \subset$ AdS$_4$ always expand until they force $F_0=0$, which takes us back to the supersymmetric case.} This would make one conclude that all non-supersymmetric AdS vacua are unstable.

In our case, such a brane would wrap a $\mathbb{R}\times S^6\subset$ AdS$_8$. There are thus only two options: a D6-brane which is a point in the internal $M_2$, and a D8-brane wrapping all of the internal $M_2$. (More generally one could consider D8/D6-bound states, but the discussion for these is the same as for a pure D8-brane.)

A D6-brane couples in fact to $F_2$; but this flux is just absent in our solution, and thus the coupling to the RR term is just absent. Only the DBI gravitational term is present, which will make such a bubble collapse, if it is created.

We next consider a D8-brane. Here we find a more fundamental problem: such a D8 would intersect transversely the O8-plane already present in the solution; this is not possible. To see why, call $\rho$ the radial direction of AdS$_8$ (in global coordinates), and say the D8 is at $\rho=\rho_0$; the O8 in our solution is of course extended along all of AdS$_8$, and sits at $z=z_0$. (While in figure \ref{fig:reg-O8} we have depicted only $z<z_0$, recall that in fact there is also a region $z>z_0$, where the graph of the functions would just be a mirror image of those for $z<z_0$.) In our original vacuum, which would exist outside the bubble, $\rho > \rho_0$, $F_0$ has values
\begin{equation}
	F_0 = \frac1{2\pi}\left\{ \begin{array}{cc}
		n_0  & \ z < z_0\,, \\
		-n_0 & \ z > z_0\,,
	\end{array}\right.
\end{equation}
because the O8 reverses the sign of $F_0$. After crossing the D8 into the $\rho < \rho_0$ region, $F_0$ should change by one unit, going to 
\begin{equation}
	F_0 \buildrel ?\over= \frac1{2\pi}\left\{ \begin{array}{cc}
		n_0 +  1& \ z < z_0\,, \\
		-n_0 + 1 & \ z > z_0\,.
	\end{array}\right.
\end{equation}
But this configuration would not be consistent with the O8 action $F_0 \to -F_0$. Thus a D8-brane bubble cannot in fact exist, and cannot destabilize our solution.\footnote{A D8-antiD8 pair would not have such a problem; moreover, the O8 projection removes the tachyon on this system and makes it stable. This stable non-BPS brane is T-dual to the seven-brane in \cite[(3.1)]{bergman-gimon-horava}) and plays a role in \cite{bergman-rodriguezgomez-zafrir}. However, it does not change $F_0$ and thus does not destabilize our solution.} 

After all this discussion, it is perhaps also worth remarking that there are no supersymmetric solutions that our solutions can decay to (unlike for the AdS$_4$ solutions of \cite[Sec.~4.1]{gaiotto-t}). Thus it is only natural that there are no decay channels. 

One last possibility would be a ``bubble of nothing''. This was shown to exist for non-supersymmetric Minkowski$_4 \times S^1$ in \cite{witten-bubble-nothing}. In that case, the surface of the bubble is a locus where the internal $S^1$ shrinks smoothly. One might imagine something like this in our case, but our $S^2$ is not round and thus cannot shrink smoothly on a locus inside AdS$_8$. One might imagine a configuration where the $S^2$ has the shape required by our vacuum at infinity, but becomes round on an interior locus. That seems unlikely to us, in particular because of the presence of the O8 at the equator. 

We now consider the solutions where D8-branes are also present. When there is a stack of D8-branes away from the O8, as in figures \ref{fig:O8D8} and \ref{fig:reg-D8-O8}, one can ask whether they are unstable against small uniform perturbations in their $z$ position in either direction. Let us investigate this in a probe approximation. The low-energy action has two terms: a DBI term $\int d^9 xe^{-\phi}\sqrt{-g}$, which generalizes the gravitational potential of a particle, and a WZ term $\int C_9$, which gives the interaction with $F_0= * F_{10}=d C_9$. For a D8 in the background of a stack of other D8-branes in flat space, the gravitational attraction would exactly balance everywhere with the repulsion given by the presence of $F_0$. 

Our curved space solution is more complicated; the gravitational attraction and the $F_0$ repulsion do not exactly balance everywhere. In fact, the sum of the two forces is proportional to $F_0 + e^{W-\phi} \left(\phi'- W' + \frac{\alpha'}\alpha\right)$ and D8 branes can only be placed at loci where this force vanishes.  Looking back at (\ref{eq:stop}), we see that this is exactly the condition we used to decide where to place our D8 stack. However, the force becomes non-zero away from the D8 stack; we find that it is positive for $z> z_\mathrm{D8}$ and negative for $z< z_\mathrm{D8}$, meaning that the D8s in the stack are in unstable equilibrium. In other words, when we move one of the D8s to the right of the stack, they are repelled by the gravitational potential of the O8, but the coupling to $F_0$ gives a stronger force which pushes them towards the O8. On the other hand, if we move one of the D8s to the left of the stack, the $F_0$ force is weaker than the gravitational repulsion of the O8, and the D8 slips off towards the regular point. 

There is no such instability for D8-branes on top of the O8-plane.  In that case, the gravitational and $F_0$ force balance on the O8, and away from it are arranged so that they lead to stable equilibrium.\footnote{In the full KK analysis, the open-string D8 degrees of freedom might interact with the supergravity fluctuations.}

% subsection stab (end)

\section{O8--D8 AdS$_d$ Solutions } % (fold)
\label{sec:adsd}

In this section, we will generalize the O8--D8 solutions of section \ref{sec:o8-d8} to arbitrary AdS$_d\times M_{10-d}$ spacetimes.  This results in a simple class of non-supersymmetric AdS$_{d}$ solutions supported only by $F_{0}$.  The compactification manifold $M_{10-d}$ will be topologically a sphere $S^{10-d}$ with SO(10-d) isometry. Parallel to our earlier AdS$_{8}$ examples, there is a $\mathbb{Z}_2$ involution and an O8-plane at its fixed point.

Explicitly, for the manifold $M_{10-d}$, we consider a fibration of a round sphere  $S^{9-d}$ (whose radius is defined by $R_{mn} = \rho g_{mn}$) over an interval identified by the coordinate $z$. We will later use regularity to fix the radius to one.
We again work in the gauge \eqref{eq:Q=W}, where our ansatz for the metric now reads
\begin{equation}\label{eq: ds2d}
ds^2_{10} = e^{2W} ds^{2}_{\text{AdS}_d}+ e^{-2W}(dz^2+ e^{2\lambda} ds^2_{S^{9-d}})\,.
\end{equation}

The equations of motions with 9-dimensional sources orthogonal to the coordinate $z$ now read:
\begin{subequations}\label{eq:eqD1}
\begin{align}
0&=8 \frac{\alpha'}\alpha (\phi+(4-d) W)' + 2 (d-8) \frac{(\alpha ')^2}{\alpha ^2}+(d-9) \left(F_0^2
   e^{\frac{(-2d+18) W}{d-9}+2\phi}+2 d   e^{\frac{(-4 d+36) W}{d-9}}\right) +\nonumber\\
&+8 \left(\phi'\right)^2 -8 (d+1) W' \phi'+8 (3 d-2) \left(W'\right)^2+2 (d-9)^2 \rho  \alpha ^{\frac{2}{d-9}} e^{\frac{4 (\phi -(d-4)
   W)}{d-9}}\,, \label{eq:eqD1-1}\\
 0&=F_0^2 e^{2 \phi }-4  -4 e^{2 W} \left(W''-\frac{\alpha'}\alpha W'\right)+\delta  \kappa ^2 \tau  e^{ W+\phi }\,, \\
 0&= (9-d) \rho  \alpha ^{\frac{2}{d-9}+1} e^{\frac{4 (3 W+\phi )}{d-9}}- \alpha'' e^{\frac{4 d W}{d-9}}- \alpha  (d+1)   e^{\frac{32 W}{d-9}}\,,\label{eq:eqD1-3}\\
 0&=(9-d)( \rho  \alpha ^{\frac{2}{d-9}} e^{\frac{4 \phi }{d-9}}+e^{\frac{20 W}{d-9}})+e^{\frac{4 (d-4) W}{d-9}} \left((10 W-2 \phi)'' + (10W-2 \phi)'\frac{\alpha'}\alpha - \frac{\alpha''}\alpha\right)~,\label{eq:eqD1-4} 
 \end{align}
\end{subequations}
where we introduced
\begin{equation}
\alpha = e^{(9-d)\lambda - 2\phi+2(d-4)W}\,,
\end{equation}
which has the property that its derivative does not jump across the brane sources, generalizing (\ref{eq:alpha}).

The first-order equation \eqref{eq:eqD1-1} expresses the expected constraints in gravitational theories, generalizing (\ref{eq:first}). (\ref{eq:eqD1}) are again invariant under the rescaling (\ref{eq:resc}). (We also have the possibility of rescaling $\lambda\to \lambda+ 2\tilde c$, $\rho\to \rho e^{4\tilde c}$; this however is just a redefinition, and does not change the solution.)

We can start analyzing the properties of the equations by looking at the behavior across the sources. Doing so we get the same conditions as in \eqref{eq:dw}.

After taking care of the behavior near the sources, we can now eliminate one of the second order equations, say \eqref{eq:eqD1-4}, and use the remaining system to look for regular solutions.

Imposing the conditions for regularity as in \eqref{eq:reg}, but without fixing the first-order coefficient of $e^{\lambda}$, we get for $d<8$ the perturbative solution
\begin{subequations}\label{eq: pertRegularD}
\begin{align}
e^{\frac{2}{9-d} \phi} &= c_1+z^2 \frac{5 F_0^2  c_1^{10-d} c_2^{9-d}}{4 \left(d^2-19 d+90\right)}+O\left(z^4\right)~,\\
e^{\frac{2}{9-d} W} &= \frac{1}{c_2}+z^2 \frac{\left(F_0^2 c_1^{9-d} c_2^{8-d}-4   c_2^{17-2 d}\right)}{4 d^2-76
   d+360}+O\left(z^4\right)~,\\
\alpha^{\frac{2}{9-d}} & = z^2 \frac{\rho c_2^{8-2 d}}{c_1^2 (8-d)}-z^4 \frac{ \rho  c_1^{-d-2} c_2^{17-4 d}   \left(c_1^9 (d-8) F_0^2 c_2^d-6 c_2^9 (d-2)  c_1^d\right)}{6 (d-10) (d-9)(d-8)}+O\left(z^6\right)\,.
\end{align}
\end{subequations}
(As in the $d=8$ case, this expansion can be pushed to high order, but given all the free parameters we have at this point the expressions become quite cumbersome very quickly.)

On this solution, $e^{2\lambda}$ behaves as:
\begin{equation}
e^{2\lambda}=z^2 \frac{\rho}{8-d}+O\left(z^4\right)\;.
\end{equation}
In order to have a regular space, we fix the value of $\rho$ such that the sphere $S^{9-d}$ has radius one:
\begin{equation}
\rho = 8-d.
\end{equation}
This choice fixes the linear coefficient in the expansion of $e^{\lambda}$ to be 1.  

The local regular solution above again depends on  two  real parameters, $c_0$ and $c_1$.  As in our previous analysis we can now numerically evolve along $z$.  By tuning the initial conditions we again find an O8 singularity.  This leaves us with a one-parameter family of solutions related by the modulus \eqref{eq:resc-10}.  This construction is possible in all $ 2\leq d< 8$ resulting in solutions qualitatively similar to those described in section \ref{sec:o8-d8} in all spacetime dimensions.

% subsection adsd-rego8 (end)

% section adsd (end)

\section*{Acknowledgements}

We thank O.~Bergman, D. Junghans, I. Klebanov, J. Maldacena, C. Nunez,  A.~Sagnotti, E. Witten, and T. Wrase for discussions. CC is supported by DOE grant de-sc0009988. GBDL and AT are supported in part by INFN. 

\appendix

\section{Equations of Motion}% (fold)
\label{app:typeIIeom}

In this appendix we summarize the equations of motion of type II string theory.

The bosonic closed-string field content consists of a metric $g_{MN}$, a dilaton $\phi$, a two-form $B$ field  with three-form field strength $H$, as well as Ramond--Ramond fields $C_{p-1}$ with field-strengths $F_p$, with $p$ even. We work with a complete set of field strengths and impose the duality relations at the same time as the equations of motion.  

The equations of motion then read
\begin{subequations}\label{eq:typeIIeom}
	\begin{align}
	R+4\nabla^{2}\phi-4\left(\nabla \phi\right)^{2}-\frac{1}{2}|H|^{2}& =  \frac12\tau \kappa^2 e^{\phi+W} \delta \,, \label{eq: typeIIeom:1}\\
	e^{-2\phi}\left(R_{MN}+2\nabla_M\nabla_N\phi-\frac{1}{2}|H|^{2}_{MN}\right)-\frac{1}{4}\sum_{p\geq1}|F_{p}|^{2}_{MN}& = \frac14 \tau \kappa^2 e^{-\phi+W} \left(g_{MN}-2 \Pi_{MN}\right)\delta \,, 
	\label{eq: typeIIeom:2}\\
	d(e^{-2\phi}*H)+\frac{1}{2}\sum_{p\geq2}F_{p-2}\wedge *F_{p}& = 0\,, \label{eq: typeIIeom:3}\\
	dF_{p}+H\wedge F_{p-2} & = - \kappa^2 \tau \delta\,,\label{eq: typeIIeom:4}\\
	*F_{p}+(-1)^{p(p+1)/2}F_{10-p} & = 0\,. \label{eq: typeIIeom:5}
	\end{align}	
\end{subequations}
We have collected the source terms on the right.
\begin{equation}\label{eq:kappa}
	\kappa^2\equiv(2\pi)^7(l_s)^8
\end{equation}
denotes Newton's constant. $\tau$ denotes the total source's tension; it is the sum $\tau = (n_{\mathrm{D}p} \tau_{\mathrm{D}p} - 8 n_{\mathrm{O}p})\tau_{\mathrm{O}p}$, where
\begin{equation}\label{eq:tau's}
	\tau_{\mathrm{D}p}=\frac1{g_s (2\pi)^p l_s^{p+1}} \, ,\qquad 
	\tau_{\mathrm{O}p}= -2^{p-5}\tau_{\mathrm{D}p}	
\end{equation}
are the D-brane and O-plane tensions; $n_{\mathrm{O}p}\in \{0,1\}$. (In the main text we work in string units $l_s =1$.) $\delta$ is locally of the form $\Pi_{m=p+1}^{10}\delta(x^m)dx^m$; the projector $\Pi$ is defined by
\begin{equation}
	\Pi_{MN}\equiv E^{\alpha \beta} \partial_\alpha x^P \partial_\beta x^Q g_{MP} g_{NQ}\,,
\end{equation}
where $\alpha$, $\beta=0,\,\ldots,\,p$ denote brane indices, $E_{MN}= g_{MN}+ B_{MN}$, its pull-back $E_{\alpha \beta}= \partial_\alpha x^M \partial_\beta x^N E_{MN}$, and its inverse is $E^{\alpha \beta}$.

% section typeIIeom (end)

\section{Other Cases} % (fold)
\label{sec:other}

We will now look for AdS$_8$ solutions in other setups. In section \ref{sub:F00} we will analyze the case $F_2\neq0$, $F_0=0$, and show that within our cohomogeneity-one ansatz there are no physical solutions. In section \ref{sub:iib} we will look at IIB, where the only solutions we found are of dubious physical significance.

\subsection{IIA, $F_2\neq 0$, $F_0=0$: no solutions} % (fold)
\label{sub:F00}

We will now look at the other branch of (\ref{eq:F0F2}), namely $F_2\neq0$, $F_0=0$, again using the cohomogeneity-one ansatz, where the metric reads (\ref{eq:gf}), with $W$ and the dilaton $\phi$ only depending on $z$. An important difference with section \ref{sec:o8-d8} is that now we have no natural candidates for sources from string theory. Indeed $F_2$ should be sourced by D6-branes, which cannot be introduced in our system without breaking the isometries of AdS$_8$.

We can parameterize 
\begin{equation}
	F_2 = f_2 dz \wedge d \theta\,.
\end{equation}
The equation of motion $d *F_2 =0$ gives, recalling
\begin{equation}
	\partial_\theta f_2 = \partial_z(e^{10W-\lambda} f_2)=0 \qquad \Rightarrow \qquad f_2 = e^{-10W+ \lambda} f_{20}\,.
\end{equation}
(The first condition is in fact part of the cohomogeneity-one ansatz.) 

We perform now the same manipulations as in the $F_0\neq 0$ case, taking again $Q=W$. We end up with the system
\begin{subequations}\label{eq:sysF2}
	\begin{align}
		\label{eq:firstF2}
	  0 & =  \frac{1}{8} f_{20}^2 e^{2 (\phi - 9 W)} + W'\left(22 W'-9 \phi'-4\frac{\alpha'}\alpha\right)
	+ \phi'\left(\phi'+\frac{\alpha'}{\alpha}\right)- 2  e^{- 4 W} \,, \\
	  0 & =  \frac{1}{4} f_{20}^2 e^{2 (\phi - 9 W)} - W'' - W'\frac{\alpha'}{\alpha} 
	-  e^{- 4 W}\,, \\
		\label{eq:lastsysF2}
	  0 & =  \frac{1}{2} f_{20}^2 e^{2 (\phi - 9 W)} - \frac{
	  \alpha''}{\alpha}- 9  e^{-4 W} \,.
	\end{align}
\end{subequations}

To make the solutions compact, one possibility is to make the circle paramerized by $\theta$ shrink at two points, leading to an $S^2$ topology. The second possibility is a periodic solution, where $z$ would be identified to itself up to a translation, leading to a $T^2$ topology.

We begin with the first possibility. The starting point is to
 look perturbatively for regular solutions, which we again impose by (\ref{eq:reg}). We get
\begin{subequations}\label{eq:pertF2}
	\begin{align}
	  W & =  \frac{\log (c_1)}{2} + z^2 \frac{(c_0^2 f_{20}^2 - 4 c_1^7)
	  }{16 c_1^9} - z^4 \frac{ (-44 c_0^2 c_1^7 f_{20}^2 + 5 c_0^4
	  f_{20}^4 + 60 c_1^{14} )}{384 c_1^{18}} + O (z^6) \\
	  \phi & =  \log (c_0) + z^2  \frac{3 c_0^2 f_{20}^2}{16 c_1^9} - z^4 \frac{
	  (-36 c_0^2 c_1^7 f_{20}^2  + 5 c_0^4 f_{20}^4)}{128 c_1^{18}} + O
	  (z^6)\,, \\
	  \alpha & =  z \frac{c_1^4}{c_0^2} + z^3
	 \frac{ \left( -18 c_1^7
	   + f_{20}^2 c_0^2\right)}{12 c_1^5 c_0^2} + \frac{1}{120} z^5  \left( 
	  \frac{18 f_{20}^2 }{c_1^7} - \frac{2 c_0^2 f_{20}^4}{c_1^{14}} +
	  \frac{27 }{c_0^2} \right) + O (z^7)  \,.
	\end{align}
\end{subequations}
We can again use this as a starting point for a numerical study, and as before we always evolve to singularities. However,  unlike in section \ref{sec:o8-d8}, we cannot interpret these as the back-reaction of some physical object. (Again there are two branches of solutions. One goes as in footnote \ref{foot:sing}; the other behaves as $e^W\sim t^{- 3 / 8}$, $e^\phi \sim t^{-21/8}$, $e^\lambda\sim t^{-5/4}$.) This is ultimately because of our observation at the beginning of this subsection, that there are no natural candidates. So the possibility of solutions with $S^2$ topology fails.

This leaves us to consider periodic solutions, with $T^2$ topology.  To exclude such solutions we observe that by taking derivatives of the first-order equation (\ref{eq:firstF2}) with the others in (\ref{eq:sysF2}), we can find a combination that reads
\begin{equation}
  \partial_z (\phi' \alpha) = \frac{3}{4} \alpha e^{- 18 W + 2 \phi} f_{20}^2\, .
\end{equation}
This equation says that $\phi' \alpha$ is monotonous; so the solution cannot be periodic. 

% subsection F00 (end)

\subsection{IIB} % (fold)
\label{sub:iib}

We will now turn to IIB. We will again use the cohomogeneity-one ansatz (\ref{eq:gf}). Again everything depends on the coordinate $z$ only. The only possible flux is now $F_1$, a one-form on $M_2$. (\ref{eq: typeIIeom:4}) tells us that it should be locally closed. The only possible source for it is an O7 or a D7 filling completely eight-dimensional spacetime, and localized in the internal $M_2$.
Since we are taking $\partial_\theta$ to be an isometry, this should be at a locus where the circle shrinks.

From the $z \theta$ component of (\ref{eq: typeIIeom:2}) we now get
\begin{equation}
  F_{1 \theta} F_{1 z} = 0 \,.
\end{equation}
A D7 or an O7 at $z=z_0$ would source $F_{1\theta}$, since $\int F_1 = \int d \theta F_{1 \theta}$ measures the object's charge. So we choose $F_{1 z} = 0$. 

We can now compute the system of ODEs. Defining $\alpha$ as in \eqref{eq:alpha}, after some manipulations we end up with the system
\begin{subequations}\label{eq:sysIIB}
\begin{align}
\frac{F_{1 \theta }^2}{8 \alpha ^2} e^{16 W-2 \phi } =&  \frac{\alpha'}{\alpha} (\phi -4 W)'+(\phi')^2-9 W' \phi'+22 (W')^2-2  e^{-4 W}\;,\\
 \frac{F_{1 \theta}}{4\alpha^2} e^{16W-2 \phi } =& \left(\alpha  W'\right)' + e^{-4 W} \;,\\
e^{4 W}  =&\, -9  \frac{\alpha}{\alpha''}\,~.
\end{align}
\end{subequations}
Again we observe that $\alpha''\leq 0$ excluding periodic solutions. The full system is invariant under the rescalings
\begin{equation}
W \to W + c_1 ,\quad \phi \to \phi-c_2,\quad \lambda \to \lambda+2c_3,\quad F_{1\theta} \to e^{2c_3+c_2-2c_1}F_{1\theta},\quad z\to e^{2c_1}z .
\end{equation}

From the system \eqref{eq:sysIIB}, we can derive a monotonicity equation:
\begin{equation}\label{eq:monIIB}
\partial_z \left(\phi' \alpha\right) = e^{16W-2\phi} \frac{F_{1\theta}^2}{\alpha} \geq 0~.
\end{equation}
Using the above we can exclude regular solutions in IIB.  Indeed, at a regular point, both  $\phi'$ and $\alpha$ vanish, and hence so the left-hand side of \eqref{eq:monIIB} also vanishes. However, if $W$ and $\phi$ remain finite, the right-hand side diverges at that point, and the equation cannot be satisfied.

That leaves us one last possibility: that the solution has two sources at two values of $z$. We can for example expand around an O7-plane. This is subtler than the O8-planes we discussed in section \ref{sec:o8-d8}. The metric for a D7-brane or O7-plane in flat space reads: 
\begin{equation}\label{eq:O7loc}
	ds^2_{10}  = H^{-1/2}(-dx_0^2 + dx_1^2 +\ldots dx_7^2)+ H^{1/2} (dx_8^2+dx_9^2)\,,
\end{equation} 
which is similar to (\ref{eq:O8loc}), but now with $H\sim a + b \log(x_8^2+x_9^2)$, with $b = -\frac{g_s n}{4 \pi}$; the string coupling is $e^\phi = g_s H^{-1}$ and $n=n_\mathrm{D7}- 4 n_\mathrm{O7}$ ($n_\mathrm{O7}\in \{0,1\}$) is the D7-charge of the object, also measured by
\begin{equation}
	n=\int F_1 = \int d\theta F_{1\theta}\,.
\end{equation}

We recover this solution by solving the analogue of \eqref{eq:sysIIB} for vanishing cosmological constant $\Lambda$ with
\begin{equation}
H\equiv e^{-4W} = c_0 - c_1 \log(z) ,\qquad e^{-\phi} = \pm \frac{F_{1\theta}}{c_1} H,\qquad \alpha = \frac{F_{1\theta}^2}{c_1^2} z\,,
\end{equation}
for which $e^\lambda = z$ and we can identify $dz^2+z^2d\theta^2  = dx^2_8 + dx^2_9$.

If $n>0$, so that the object has positive total tension, there is an excluded region for large enough $dx_8^2+dx_9^2$, where the metric doesn't make sense; if $n<0$ and negative total tension, the excluded region is at small $x_8^2+x_9^2$. The boundary of this region is $x_8^2+x_9^2=1$ for $a=0$.

The strange behavior at small distance is in fact cured by non-perturbative physics, as can be found using F-theory \cite{greene-shapere-vafa-yau} (see for example \cite[Sec.~3]{denef-leshouches} for a review):
\begin{equation}
	H= e^{- \phi} \frac{|\eta(\tau)|^4}{\Delta^{1/6}(\tau)}
\end{equation}
where $e^\phi$ is determined by $\tau= C_0 + i e^{-\phi}$ and by inverting $j(\tau)= \frac4 \Delta (24 f)^3$; $j$ is the modular invariant function of the fundamental region under $\mathrm{SL}(2,\mathbb{Z})$, $\Delta= 27 g^2 + 4 f^3$ is the discriminant, $f$ and $g$ are the functions of $\tau$ defined by the Weierstrass equation $y^2= x^3 + f x + g$ for a torus of modular parameter $\tau$. Finally, $\tau$ is a holomorphic function of $u \equiv x_8+i x_9$. A D7-brane is realized by $j(\tau(u))= \frac1u$. An O7-plane turns out to be \cite{sen-O7} a bound state of two $\mathrm{SL}(2,\mathbb{Z})$ duals of D7-branes; the excluded region is now no longer present.

Within supergravity, however, we can only use the description (\ref{eq:O7loc}). We cannot expand around the center $x_8=x_9=0$, where the metric does not make sense, but we can expand around the boundary of the excluded region. (There are by now many examples where a local metric has been successfully identified with an O-plane by comparing its behavior near the excluded region; for example, AdS$_7$ solutions include O6-planes \cite{afrt}, and holography works well in their presence \cite{apruzzi-fazzi}.) In this spirit we impose 
\begin{equation}\label{eq:behaviorO7}
	e^{-4W} \sim t \, ,\qquad e^{-\phi} \sim t \, ,\qquad \alpha \sim \text{const}. \, ,\qquad t\equiv |z-z_0|\,.
\end{equation}
One can indeed find a perturbative solution with this behavior.
The first orders of this expansion are
\begin{subequations}\label{eq:locO7}
\begin{align}
e^{-\phi} &=\frac{\alpha _0 h_1^2 t}{F_{1\theta }}+\varphi _2 t^2+t^3\frac{ \left(-12 \alpha _1 h_1^2 \varphi _2 F_{1\theta }+20 \varphi _2^2 F_{1\theta }^2+21 \alpha _1^2 h_1^4\right)}{96 \alpha _0 h_1^2 F_{1\theta}}+O\left(t^4\right)\,,\\
e^{-4W} &= h_1 t-t^2\frac{ \left(\alpha _1 h_1^2-2 \varphi _2 F_{1\theta }\right)}{4 \left(\alpha _0 h_1\right)}+t^3\frac{ \left(9 \alpha _1^2 h_1^4-4 \varphi _2^2 F_{1\theta }^2\right)}{24 \alpha _0^2
   h_1^3}+O\left(t^4\right)\,,\\
\alpha &=\alpha _0+\alpha _1 t-\frac{3}{2} \alpha _0 h_1   t^3-  t^4 \left(\frac{3 \varphi _2 F_{1\theta }}{8 h_1}+\frac{9 \alpha _1 h_1}{16}\right)+O\left(t^5\right)\,.
\end{align}
\end{subequations} 
(Again the expansion can be pushed to high orders; see footnote \ref{foot:pert}.) By choosing some value for the free parameters and starting the numerical evolution, we get two kinds of numerical solutions. In one case we have a non-physical divergence. In the other case,
the solution is attracted back to an endpoint with the behavior \eqref{eq:behaviorO7}. This looks like another O7-plane on the other side; but Gauss's law implies that it has in fact negative tension and positive charge. While such an ``anti-orientifold'' does exist, it is not entirely clear that it makes sense to combine it with an ordinary one.

% subsection iib (end)                 

% section other (end)

\bibliography{at}

\providecommand{\href}[2]{#2}\begin{thebibliography}{10}

\bibitem{Witten:1995zh}
E.~Witten, ``{Some comments on string dynamics},'' in {\em {Future perspectives
  in string theory. Proceedings, Conference, Strings'95, Los Angeles, USA,
  March 13-18, 1995}}, pp.~501--523.
\newblock 1995.
\newblock
\href{http://arXiv.org/abs/hep-th/9507121}{{\tt hep-th/9507121}}.
\newblock
%%CITATION = HEP-TH/9507121;%%.

\bibitem{Strominger:1995ac}
A.~Strominger, ``{Open p-branes},'' {\em Phys. Lett.} {\bf B383} (1996) 44--47,
\href{http://arXiv.org/abs/hep-th/9512059}{{\tt hep-th/9512059}}.
%%CITATION = HEP-TH/9512059;%%.

\bibitem{Witten:1995em}
E.~Witten, ``{Five-branes and M theory on an orbifold},'' {\em Nucl. Phys.}
  {\bf B463} (1996) 383--397, \href{http://arXiv.org/abs/hep-th/9512219}{{\tt
  hep-th/9512219}}.
[,172(1995)].
%%CITATION = HEP-TH/9512219;%%.

\bibitem{seiberg-5d}
N.~Seiberg, ``{Five-dimensional SUSY field theories, nontrivial fixed points
  and string dynamics},'' {\em Phys.Lett.} {\bf B388} (1996) 753--760,
\href{http://arXiv.org/abs/hep-th/9608111}{{\tt hep-th/9608111}}.
%%CITATION = HEP-TH/9608111;%%.

\bibitem{afrt}
F.~Apruzzi, M.~Fazzi, D.~Rosa, and A.~Tomasiello, ``{All AdS$_7$ solutions of
  type II supergravity},'' {\em JHEP} {\bf 1404} (2014) 064,
\href{http://arXiv.org/abs/1309.2949}{{\tt 1309.2949}}.
%%CITATION = ARXIV:1309.2949;%%.

\bibitem{10letter}
F.~Apruzzi, M.~Fazzi, A.~Passias, A.~Rota, and A.~Tomasiello,
  ``{Six-Dimensional Superconformal Theories and their Compactifications from
  Type IIA Supergravity},'' {\em Phys. Rev. Lett.} {\bf 115} (2015), no.~6,
  061601,
\href{http://arXiv.org/abs/1502.06616}{{\tt 1502.06616}}.
%%CITATION = ARXIV:1502.06616;%%.

\bibitem{cremonesi-t}
S.~Cremonesi and A.~Tomasiello, ``{6d holographic anomaly match as a continuum
  limit},'' {\em JHEP} {\bf 05} (2016) 031,
\href{http://arXiv.org/abs/1512.02225}{{\tt 1512.02225}}.
%%CITATION = ARXIV:1512.02225;%%.

\bibitem{dhoker-gutperle-karch-uhlemann}
E.~D'Hoker, M.~Gutperle, A.~Karch, and C.~F. Uhlemann, ``{Warped $AdS_6\times
  S^2$ in Type IIB supergravity I: Local solutions},'' {\em JHEP} {\bf 08}
  (2016) 046,
\href{http://arXiv.org/abs/1606.01254}{{\tt 1606.01254}}.
%%CITATION = ARXIV:1606.01254;%%.

\bibitem{dhoker-gutperle-uhlemann}
E.~D'Hoker, M.~Gutperle, and C.~F. Uhlemann, ``{Holographic duals for
  five-dimensional superconformal quantum field theories},''
\href{http://arXiv.org/abs/1611.09411}{{\tt 1611.09411}}.
%%CITATION = ARXIV:1611.09411;%%.

\bibitem{heckman-morrison-vafa}
J.~J. Heckman, D.~R. Morrison, and C.~Vafa, ``{On the Classification of 6D
  SCFTs and Generalized ADE Orbifolds},'' {\em JHEP} {\bf 05} (2014) 028,
  \href{http://arXiv.org/abs/1312.5746}{{\tt 1312.5746}}.
[Erratum: JHEP06,017(2015)].
%%CITATION = ARXIV:1312.5746;%%.

\bibitem{heckman-morrison-rudelius-vafa}
J.~J. Heckman, D.~R. Morrison, T.~Rudelius, and C.~Vafa, ``{Atomic
  Classification of 6D SCFTs},'' {\em Fortsch. Phys.} {\bf 63} (2015) 468--530,
\href{http://arXiv.org/abs/1502.05405}{{\tt 1502.05405}}.
%%CITATION = ARXIV:1502.05405;%%.

\bibitem{osty-a6}
K.~Ohmori, H.~Shimizu, Y.~Tachikawa, and K.~Yonekura, ``{Anomaly polynomial of
  general 6d SCFTs},'' {\em PTEP} {\bf 2014} (2014), no.~10, 103B07,
\href{http://arXiv.org/abs/1408.5572}{{\tt 1408.5572}}.
%%CITATION = ARXIV:1408.5572;%%.

\bibitem{intriligator-a6}
K.~Intriligator, ``{6d, $ \mathcal{N}=\left(1,\;0\right) $ Coulomb branch
  anomaly matching},'' {\em JHEP} {\bf 10} (2014) 162,
\href{http://arXiv.org/abs/1408.6745}{{\tt 1408.6745}}.
%%CITATION = ARXIV:1408.6745;%%.

\bibitem{cordova-dumitrescu-yin}
C.~C\'{o}rdova, T.~T. Dumitrescu, and X.~Yin, ``{Higher Derivative Terms,
  Toroidal Compactification, and Weyl Anomalies in Six-Dimensional $(2,0)$
  Theories},''
\href{http://arXiv.org/abs/1505.03850}{{\tt 1505.03850}}.
%%CITATION = ARXIV:1505.03850;%%.

\bibitem{cordova-dumitrescu-intriligator-a6}
C.~C\'{o}rdova, T.~T. Dumitrescu, and K.~Intriligator, ``{Anomalies,
  Renormalization Group Flows, and the $a$-Theorem in Six-Dimensional
  $\left(1,\;0\right)$ Theories},''
\href{http://arXiv.org/abs/1506.03807}{{\tt 1506.03807}}.
%%CITATION = ARXIV:1506.03807;%%.

\bibitem{nahm}
W.~Nahm, ``{Supersymmetries and their Representations},'' {\em Nucl.Phys.} {\bf
  B135} (1978)
149.
%%CITATION = NUPHA,B135,149;%%.

\bibitem{minwalla}
S.~Minwalla, ``{Restrictions imposed by superconformal invariance on quantum
  field theories},'' {\em Adv. Theor. Math. Phys.} {\bf 2} (1998) 783--851,
\href{http://arXiv.org/abs/hep-th/9712074}{{\tt hep-th/9712074}}.
%%CITATION = HEP-TH/9712074;%%.

\bibitem{heemskerk-penedones-polchinski-sully}
I.~Heemskerk, J.~Penedones, J.~Polchinski, and J.~Sully, ``{Holography from
  Conformal Field Theory},'' {\em JHEP} {\bf 10} (2009) 079,
\href{http://arXiv.org/abs/0907.0151}{{\tt 0907.0151}}.
%%CITATION = ARXIV:0907.0151;%%.

\bibitem{fitzpatrick-kaplan-poland}
A.~L. Fitzpatrick, J.~Kaplan, and D.~Poland, ``{Conformal Blocks in the Large
  $D$ Limit},'' {\em JHEP} {\bf 08} (2013) 107,
\href{http://arXiv.org/abs/1305.0004}{{\tt 1305.0004}}.
%%CITATION = ARXIV:1305.0004;%%.

\bibitem{giombi-perlmutter}
S.~Giombi and E.~Perlmutter, ``{Double-Trace Flows and the Swampland},'' {\em
  JHEP} {\bf 03} (2018) 026,
\href{http://arXiv.org/abs/1709.09159}{{\tt 1709.09159}}.
%%CITATION = ARXIV:1709.09159;%%.

\bibitem{brandhuber-oz}
A.~Brandhuber and Y.~Oz, ``{The D4--D8 brane system and five-dimensional fixed
  points},'' {\em Phys.Lett.} {\bf B460} (1999) 307--312,
\href{http://arXiv.org/abs/hep-th/9905148}{{\tt hep-th/9905148}}.
%%CITATION = HEP-TH/9905148;%%.

\bibitem{bah-passias-t}
I.~Bah, A.~Passias, and A.~Tomasiello, ``{AdS$_{5}$ compactifications with
  punctures in massive IIA supergravity},'' {\em JHEP} {\bf 11} (2017) 050,
\href{http://arXiv.org/abs/1704.07389}{{\tt 1704.07389}}.
%%CITATION = ARXIV:1704.07389;%%.

\bibitem{dibitetto-lomonaco-passias-petri-t}
G.~Dibitetto, G.~Lo~Monaco, A.~Passias, N.~Petri, and A.~Tomasiello, ``{AdS$_3$
  solutions with exceptional supersymmetry},''
\href{http://arXiv.org/abs/1807.06602}{{\tt 1807.06602}}.
%%CITATION = ARXIV:1807.06602;%%.

\bibitem{dine-seiberg}
M.~Dine and N.~Seiberg, ``{Is the Superstring Weakly Coupled?},'' {\em Phys.
  Lett.} {\bf 162B} (1985)
299--302.
%%CITATION = PHLTA,162B,299;%%.

\bibitem{ooguri-vafa-nonsusy-ads}
H.~Ooguri and C.~Vafa, ``{Non-supersymmetric AdS and the Swampland},'' {\em
  Adv. Theor. Math. Phys.} {\bf 21} (2017) 1787--1801,
\href{http://arXiv.org/abs/1610.01533}{{\tt 1610.01533}}.
%%CITATION = ARXIV:1610.01533;%%.

\bibitem{narayan-trivedi}
P.~Narayan and S.~P. Trivedi, ``{On The Stability Of Non-Supersymmetric AdS
  Vacua},'' {\em JHEP} {\bf 07} (2010) 089,
\href{http://arXiv.org/abs/1002.4498}{{\tt 1002.4498}}.
%%CITATION = ARXIV:1002.4498;%%.

\bibitem{gaiotto-t}
D.~Gaiotto and A.~Tomasiello, ``{The gauge dual of Romans mass},'' {\em JHEP}
  {\bf 01} (2010) 015,
\href{http://arXiv.org/abs/0901.0969}{{\tt 0901.0969}}.
%%CITATION = 0901.0969;%%.

\bibitem{cordova-deluca-t-kk}
C.~C\'{o}rdova, G.~B. De~Luca, and A.~Tomasiello, ``{Kaluza--Klein reduction
  around a warped AdS$_8$ solution}.'' Work in progress.

\bibitem{rota-t}
A.~Rota and A.~Tomasiello, ``{AdS$\_{4}$ compactifications of AdS$\_{7}$
  solutions in type II supergravity},'' {\em JHEP} {\bf 07} (2015) 076,
\href{http://arXiv.org/abs/1502.06622}{{\tt 1502.06622}}.
%%CITATION = ARXIV:1502.06622;%%.

\bibitem{policastro-tsimpis}
G.~Policastro and D.~Tsimpis, ``{$R^4$, purified},'' {\em Class. Quant. Grav.}
  {\bf 23} (2006) 4753--4780,
\href{http://arXiv.org/abs/hep-th/0603165}{{\tt hep-th/0603165}}.
%%CITATION = HEP-TH/0603165;%%.

\bibitem{liu-minasian}
J.~T. Liu and R.~Minasian, ``{Higher-derivative couplings in string theory:
  dualities and the B-field},'' {\em Nucl. Phys.} {\bf B874} (2013) 413--470,
\href{http://arXiv.org/abs/1304.3137}{{\tt 1304.3137}}.
%%CITATION = ARXIV:1304.3137;%%.

\bibitem{arkanihamed-motl-nicolis-vafa}
N.~Arkani-Hamed, L.~Motl, A.~Nicolis, and C.~Vafa, ``{The String landscape,
  black holes and gravity as the weakest force},'' {\em JHEP} {\bf 06} (2007)
  060,
\href{http://arXiv.org/abs/hep-th/0601001}{{\tt hep-th/0601001}}.
%%CITATION = HEP-TH/0601001;%%.

\bibitem{bergman-gimon-horava}
O.~Bergman, E.~G. Gimon, and P.~Horava, ``{Brane transfer operations and
  T-duality of nonBPS states},'' {\em JHEP} {\bf 04} (1999) 010,
\href{http://arXiv.org/abs/hep-th/9902160}{{\tt hep-th/9902160}}.
%%CITATION = HEP-TH/9902160;%%.

\bibitem{bergman-rodriguezgomez-zafrir}
O.~Bergman, D.~{Rodr\'iguez--G\'omez}, and G.~Zafrir, ``{Discrete $\theta$ and
  the 5d superconformal index},'' {\em JHEP} {\bf 01} (2014) 079,
\href{http://arXiv.org/abs/1310.2150}{{\tt 1310.2150}}.
%%CITATION = ARXIV:1310.2150;%%.

\bibitem{witten-bubble-nothing}
E.~Witten, ``{Instability of the Kaluza--Klein Vacuum},'' {\em Nucl. Phys.}
  {\bf B195} (1982)
481--492.
%%CITATION = NUPHA,B195,481;%%.

\bibitem{greene-shapere-vafa-yau}
B.~R. Greene, A.~D. Shapere, C.~Vafa, and S.-T. Yau, ``{Stringy Cosmic Strings
  and Noncompact Calabi--Yau Manifolds},'' {\em Nucl. Phys.} {\bf B337} (1990)
1--36.
%%CITATION = NUPHA,B337,1;%%.

\bibitem{denef-leshouches}
F.~Denef, ``{Les Houches Lectures on Constructing String Vacua},'' {\em Les
  Houches} {\bf 87} (2008) 483--610,
\href{http://arXiv.org/abs/0803.1194}{{\tt 0803.1194}}.
%%CITATION = ARXIV:0803.1194;%%.

\bibitem{sen-O7}
A.~Sen, ``{F-theory and Orientifolds},'' {\em Nucl.Phys.} {\bf B475} (1996)
  562--578,
\href{http://arXiv.org/abs/hep-th/9605150}{{\tt hep-th/9605150}}.
%%CITATION = HEP-TH/9605150;%%.

\bibitem{apruzzi-fazzi}
F.~Apruzzi and M.~Fazzi, ``{AdS$_{7}$/CFT$_{6}$ with orientifolds},'' {\em
  JHEP} {\bf 01} (2018) 124,
\href{http://arXiv.org/abs/1712.03235}{{\tt 1712.03235}}.
%%CITATION = ARXIV:1712.03235;%%.

\end{thebibliography}
\bibliographystyle{at}

\end{document}